\documentclass[12pt]{article}
\usepackage{graphicx}

\makeatother

\newcommand{\hateq}[0]{\mathrel{\widehat\mathalpha{=}}}

\newcommand{\dual}{{}^\star}
\newcommand{\emF}{{\mathbf{F}}}

\newcommand{\rh}{r_{\mathrm{H}}}
\newcommand{\qh}{Q_{\mathrm{H}}}

\newcommand{\kh}{\kappa_{\mathrm{H}}}
\newcommand{\ph}{\Phi_{\mathrm{H}}}
\newcommand{\ah}{a_{\mathrm{H}}}
\newcommand{\dr}{\Delta (r)}

\def\l{\ell}
\def\bar{\overline}
\def\={\hateq}

\makeatletter

\@addtoreset{equation}{section}

\newcommand{\address}[1]{\vbox{\let\\=\cr \normalsize \vskip 1em
  \lineskip\normallineskip \halign{\hfil##\hfil\crcr#1\crcr}}}

\begin{document}

\title{Distorted Black Holes with Charge}
\date{}
\author {Stephen Fairhurst\thanks{E-mail:
fairhurs@phys.psu.edu} and Badri Krishnan\thanks{E-mail:
krishnan@phys.psu.edu}
\\
\address{Center for Gravitational Physics and Geometry \\
Department of Physics, The Pennsylvania State University \\ University
Park, PA 16802, USA}}

{\maketitle}

\begin{abstract}
We present new solutions to the Einstein--Maxwell equations
representing a class of charged distorted black holes.  These
solutions are static--axisymmetric and are generalizations of the
distorted black hole solutions studied by Geroch and Hartle. 
Physically, they represent a charged black hole distorted by external
matter fields.  We discuss the zeroth and first law for these black
holes.  The first law is proved in two different forms, one motivated
by the isolated horizon framework and the other using normalizations
at infinity.
\end{abstract}


\section{Introduction}
\label{s1}

Among the most intriguing results in general relativity are the black
hole uniqueness theorems (see e.g.\cite{mh}).  For Einstein--Maxwell
theory in static spacetimes, the theorem ensures that the
Reissner--Nordstr\"om metric is the unique black hole solution with a
regular event horizon and static, asymptotically flat domain of outer
communications.  However, there are some obvious situations where the
hypotheses of this theorem and hence its conclusions do not hold. 
First of all, uniqueness may fail if we consider matter other than
Maxwell fields coupled to gravity.  For example, it is now well known
that black holes can have Yang-Mills ``hair'' \cite{vg,bm}, i.e. there
is no such uniqueness theorem for Einstein-Yang-Mills theory.  Second,
there are solutions in Einstein--Maxwell theory which are
asymptotically flat but do not have a regular event horizon.  For
example, in the C-metric solutions \cite{kw,ad} the horizon contains a
nodal singularity.  Finally, there are static solutions which have a
regular event horizon but are not asymptotically flat.  They can
nonetheless be physically interesting as descriptions of the near
horizon geometry of isolated black holes which are distorted by the
presence of far away matter \cite{gh,ms}.  There are also solutions
which describe a black hole immersed in a magnetic field \cite{fje}. 
However, in all these cases, the distorted black holes themselves do
not carry any charge.  The purpose of this paper is to present the
first family of solutions to the Einstein--Maxwell equations
representing distorted charged black holes.  Although these charged
black holes are not of direct astrophysical interest, they provide an
instructive testing ground for numerous conceptual issues related to
black hole mechanics and thermodynamics.

The solutions presented in this paper are a natural extension of the
uncharged distorted black holes introduced in \cite{gh,ms}.  In
vacuum, static axi-symmetric spacetimes, Einstein's equations reduce
to Laplace's equation on flat space.  Since this equation is linear,
distorted Schwarzschild black holes can be obtained by adding an
appropriate distortion function to the Schwarzschild solution.  This
strategy cannot be extended to Einstein--Maxwell theory because we
obtain a non-linear set of coupled partial differential equations. 
However, there exists a remarkable mapping \cite{w,gha} which takes a
static, axi-symmetric vacuum solution to a non-trivial class of static
solutions in Einstein--Maxwell theory.  In particular, the
Schwarzschild family is mapped to the Reissner--Nordstr\"om family
under this transformation.  In order to obtain a class of distorted,
charged black holes, we apply this transformation to the distorted
Schwarzschild spacetimes of \cite{gh}.  The resulting class of
solutions are static and have a regular event horizon but, as in the
uncharged case, they are not asymptotically flat.  Therefore, fall
outside the scope of the uniqueness theorems.  Furthermore, we cannot
introduce the notion of null infinity and consequently, the standard
concept of an event horizon is not applicable.  Nonetheless, these
solutions can be interpreted as representing black holes via two
arguments.  First, they do admit locally defined isolated horizons
\cite{abf,afk}.  Alternatively, the solutions may be extended as
asymptotically flat spacetimes by adding matter far away from the
isolated horizon; this allows us to identify the isolated horizon as
an event horizon.  The additional matter necessarily lies outside the
scope of Einstein--Maxwell theory so that, once again, the uniqueness
theorems are not applicable.

The black holes presented in this paper obey the zeroth and first laws
of black hole mechanics.  While the zeroth law is unambiguous, we
shall discuss two derivations of the first law.  The first is based
upon the isolated horizon framework and is intrinsically local to the
horizon while the second relies on the global structure of spacetime. 
For static solutions in Einstein--Maxwell theory, the first laws
obtained by the two methods are identical.  However, even though the
solutions presented in this paper are static, we obtain two quite
different versions of the first law.  The difference arises because
the spacetime can only be extended to be asymptotically flat in the
presence of matter fields.

In standard treatments of the first law, $\delta M = \frac{1}{8\pi G}
\kappa \delta a + \Phi \delta Q$, one typically considers globally
static (or stationary) electrovac spacetimes and small departures
therefrom.  It is straightforward to calculate the area $a$ and charge
$Q$ of the horizon.  However, the surface gravity $\kappa$ and
electric potential $\Phi$ of the horizon can only be defined
unambiguously once the Killing field has been appropriately normalized
at infinity.  The mass of the black hole is taken to be the ADM mass
of the spacetime and the variation of the mass leads to the first law. 
In this framework, the area and charge of the black hole are defined
locally while the mass, surface gravity and electric potential refer
to infinity.

The global form of the first law for distorted charged black holes is
a generalization to Einstein--Maxwell theory of the first law found by
Geroch and Hartle \cite{gh} in the uncharged case.  This first law
contains extra terms which are interpreted as work terms due to the
matter fields surrounding the black hole.  However, the interpretation
of the extra terms in the first law is only heuristic.  Furthermore,
although the black hole mass appearing in this first law satisfies a
Smarr formula, it is not the ADM mass of the spacetime.

The isolated horizon framework on the other hand, only refers to
quantities defined intrinsically on the horizon (see \cite{abf,afk}
for detailed definitions and discussions).  In particular, there could
be gravitational or electromagnetic radiation in an arbitrary
neighbourhood of the horizon so long as none of it crosses the
horizon.  This framework is applicable to the distorted, charged
solutions presented here provided the additional matter fields admit a
Hamiltonian description and vanish in a neighbourhood of the horizon. 
The first law arises as a necessary and sufficient condition for time
evolution to be Hamiltonian \cite{afk}.  Thus, we obtain many
different first laws, one for each allowed time evolution.  The
isolated horizon first law takes the standard form, i.e. there are no
additional work terms.  Furthermore, by appealing to the
Reissner--Nordstr\"om solutions we can select a preferred notion of
time evolution and hence a canonical choice of black hole energy.

In section \ref{s2} we review the Weyl formalism which describes all
static, axisymmetric solutions, both in vacuum and Einstein--Maxwell
theory.  We also describe a technique \cite{gha} which allows one to obtain a
static axisymmetric solution to the Einstein--Maxwell equations from a
vacuum one.  In section \ref{s3} we describe the distorted black hole
solutions.  First we discuss the uncharged case.  Then, applying the
techniques introduced in section \ref{s2}, we construct distorted black
hole solutions with charge and discuss their properties.  Finally, in
section \ref{s4} we discuss the zeroth and first laws of black hole
mechanics.

\section{Weyl Solutions}
\label{s2}

It is well known that all static axisymmetric solutions to Einstein's
equations can be expressed in Weyl form,
\begin{equation} \label{weylmetric}
ds^{2}=-e^{2\psi}\, dt^{2} + e^{2(\gamma-\psi)}(d\rho^{2}+dz^{2}) +
e^{-2\psi}\rho^{2} \, d\phi^{2}
\end{equation}
where $\psi$ and $\gamma$ are functions of $\rho$ and $z$ only.  Let
$\Sigma$ be the three dimensional Riemannian manifold orthogonal to
the static Killing field.  Then $(\rho,z,\phi)$ are coordinates on
$\Sigma$.  In vacuum, the field equation for $\psi$ is:
\begin{equation}\label{psi}
\psi_{,\rho\rho} + \rho^{-1}\psi_{,\rho} + \psi_{,zz}  = 0\, .
\end{equation}
Let us introduce a fictitious flat metric, $h_{ab}$, on $\Sigma$ given
by $ds^{2}_{h}=d\rho^{2} + dz^{2} + \rho^{2}\, d\phi^{2}$.  Then
(\ref{psi}) is simply the Laplace equation for $\psi$ in $\Sigma$ with
respect to the flat metric $h_{ab}$.  Laplace's equation is
particularly simple to solve and has the added advantage of being
linear --- if $\psi_1$ and $\psi_2$ are solutions, then so is $\psi_1
+ \psi_2$.  This linearity will be used critically in obtaining
distorted black hole solutions.  Once a solution for $\psi$ has been
found, the second metric function $\gamma$ is obtained by simple
integration of the remaining field equations:
\begin{equation}\label{gamma}
\gamma_{,\rho} = \rho[ (\psi_{,\rho})^{2} - (\psi_{,z})^{2}] \quad {\rm
and} \quad \gamma_{,z} = 2\rho\psi_{,\rho}\psi_{,z} \,.
\end{equation}
The integrability of these equations is ensured by (\ref{psi}). We
impose the boundary condition that $\gamma = 0$ on the $\rho = 0$ axis
at all points where $\psi$ is nonsingular; this ensures the
circumference of a circle with radius $r$, centered on the $z-$axis
away from material singularities will be $2\pi r$ for $r \rightarrow
0$.  This condition also serves to fix the constant freedom in
$\gamma$. In fact, if $\gamma =0$ at any point $p$ on the $z-$axis,
then (\ref{gamma}) ensures that it will vanish at all points of the
axis which are connected to $p$. Finally, it is clear from the form of
the metric (\ref{weylmetric}) that a solution will be asymptotically
flat (and the static Killing field $\partial/\partial t$ be unit
timelike at infinity) if and only if $\psi$ and $\gamma$ tend to zero
at infinity.

Let us now consider static axisymmetric solutions to the
Einstein--Maxwell equations.  By performing a gauge transformation, the
electromagnetic potential can always be cast in the form
\[
A=\Phi dt + \beta d\phi \, ,
\]
where $\Phi$ can be thought of as the electromagnetic potential and
$\beta$ as the magnetic potential.  For the remainder of the paper, we
shall be interested only in electric fields and will set $\beta = 0$. 
However, our results can be generalized to the case of non-vanishing
magnetic field, for details see \cite{zp,gha}.  Alternatively,
solutions with non-vanishing magnetic fields can be obtained by
performing a duality rotation $F \rightarrow \dual F$ (and keeping the
spacetime metric unchanged) on a purely electric solution.

In the Einstein--Maxwell case, the metric takes the same form as before
(\ref{weylmetric}) but the field equations are now expressed in terms
of the two metric functions $\psi$, $\gamma$ and the electromagnetic
potential $\Phi$. The resulting equations are (setting $G = 1$):
\begin{eqnarray}\label{einsmax}
    &&\psi_{,\rho\rho} + \rho^{-1}\psi_{,\rho} + \psi_{,zz}  =
        e^{-2\psi}(\Phi_{,\rho}^{2} + \Phi_{,z}^{2}) \nonumber \\
    &&(\rho\Phi_{,\rho}e^{-2\psi})_{,\rho} + (\rho\Phi_{,z}e^{-2\psi})_{,z}
        = 0 \nonumber \\
    &&\gamma_{,\rho} = \rho[ (\psi_{,\rho})^{2} - (\psi_{,z})^{2}
        - e^{-2\psi}(\Phi_{,\rho}^{2} - \Phi_{,z}^{2})]  \\
    &&\gamma_{,z} = 2\rho[\psi_{,\rho}\psi_{,z}
        - e^{-2\psi}\Phi_{,\rho}\Phi_{,z}] \, .  \nonumber
\end{eqnarray}
As in the vacuum case, one can solve for $\psi$ and $\Phi$ and then
integrate the last two equations to find $\gamma$.  However, the
equation for $\psi$ now has source terms and is no longer linear in
$\psi$.  Thus, even if there were a simple method of obtaining
solutions for $\psi$ and $\Phi$, the nonlinearity of the equations
would prevent us from ``distorting'' the known black hole solutions. 
There is, however, a simple method of obtaining a class of static,
axisymmetric solutions to the Einstein--Maxwell equations from the
corresponding {\textit{vacuum}} solutions.  This class contains all
electrovac solutions in which there is a functional relationship
between the gravitational potential $\psi$ and the electromagnetic
potential $\Phi$, i.e $\psi = \psi(\Phi)$.  Although this restricts
the class of spacetimes under consideration, we shall see later that
the Reissner--Nordstr\"om solutions are permitted.  Therefore, this
class will be of interest for obtaining distorted black holes with
charge.

Let us now describe how to obtain a solution to the Einstein--Maxwell
equations from a vacuum solution \cite{gha,w}.  Given the quadruple
$(\bar{\psi},\bar{\gamma},C,v)$ where $\bar{\psi}$ and $\bar{\gamma}$
satisfy the vacuum Einstein equations (\ref{psi}), (\ref{gamma}) and
$C$ and $v$ are constants, we shall construct a solution
$(\psi,\gamma,\Phi)$ to the Einstein--Maxwell equations
(\ref{einsmax}).  First, the potential $\psi$ is given in terms of
$\bar{\psi}$, $C$ and $v$ as:
\begin{equation} \label{eqforpsi}
       e^{-\psi} = \frac{e^{-v}}{2} \left[ \left(1 +
       C{(C^{2}-1)}^{-1/2} \right) e^{-\bar{\psi}} + \left(1 -
       C{(C^{2}-1)}^{-1/2} \right) e^{\bar{\psi}} \, \right] \, .
\end{equation}
Next, we turn our attention to the electromagnetic potential.  If one
assumes a functional relationship between $\psi$ and $\Phi$, it follows
from the first two equations in (\ref{einsmax}) that
\begin{equation}
    \frac{d^2 (e^{2\psi})}{d \Phi^2} = 2 \, .
\end{equation}
It is simple to integrate this equation twice to obtain an expression
for $\Phi$,
\begin{equation} \label{eqforphi}
    e^{2\psi} = e^{2v} - 2e^{v}C\Phi + \Phi^{2} \, .
\end{equation}
The integration constants have been chosen such that, with $\psi$
given by (\ref{eqforpsi}), the first two Einstein--Maxwell equations
(\ref{einsmax}) hold.  In an asymptotically flat spacetime, we require
that $\psi \rightarrow 0$ at infinity.  Therefore, in such spacetimes
$v$ must be set to zero in order that $\Phi$ vanishes at infinity.

Finally, we must specify the form of $\gamma$.  Remarkably, given
$\psi$ and $\Phi$ from (\ref{eqforpsi}) and (\ref{eqforphi})
respectively, the function $\bar{\gamma}$ satisfies the last two
Einstein--Maxwell equations (\ref{einsmax}).  Therefore, we set
\begin{equation}\label{eqforgamma}
    \gamma = \bar{\gamma}.
\end{equation}
It is straightforward to verify that $(\psi,\gamma,\Phi)$ is indeed a
solution to the Einstein--Maxwell equations (\ref{einsmax}). 
Furthermore, the function $\bar{\psi}$ from which $\psi$ and $\Phi$
are obtained satisfies the Laplace equation.  Therefore, we have
reduced the task of solving (\ref{einsmax}) to solving the vacuum
Laplace equation for $\bar{\psi}$ and a first order equation for
$\bar{\gamma}$; all other manipulations are purely algebraic.

\section{Distorted Black Holes}
\label{s3}

In this section we shall obtain distorted charged black hole
solutions.  These solutions generalize the previously known vacuum
distorted black hole solutions of \cite{gh,ms} to the case of
non-vanishing electric fields.  For completeness, we begin with a
review of the uncharged solutions.  We then describe how these
solutions can be generalized to Einstein--Maxwell theory using the
techniques outlined in section \ref{s2}.  Finally, we discuss some
interesting properties of these solutions.  In section \ref{s4}, we
shall focus on the thermodynamics of these black holes.

\subsection{Distorted Schwarzschild Solutions}
\label{s3.1}

The exterior region of the Schwarzschild solution is static and has an
axial Killing vector.  Therefore it is a vacuum Weyl solution and
corresponds to a specific choice of $\bar{\psi}$.%
\footnote{In this subsection, we
shall denote the functions corresponding to the Schwarzschild and
distorted Schwarzschild solutions with an overbar i.e. $\bar{\psi}$
and $\bar{\gamma}$ to distinguish them from the distorted charged
black holes introduced in section \ref{s3.2}.}
Interestingly, $\bar{\psi}$ for a Schwarzschild black hole of mass $A$
is just the flat space Newtonian potential due to a rod of length $2A$
and mass $A$ placed symmetrically on the $\rho = 0$ axis.  The
function $\bar{\gamma}$ is then determined uniquely from $\bar{\psi}$
by (\ref{gamma}) whence we obtain,
\begin{equation}\label{schwarzfn}
    \bar{\psi}_{\rm S} := \frac{1}{2}\ln \left(\frac{L-A}{L+A}\right)
    \quad {\rm and} \quad
    \bar{\gamma}_{\rm S} := \frac{1}{2}
        \ln \left(\frac{L^{2} - A^{2}}{L^{2}-\eta^{2}}\right)\, ,
\end{equation}
where $L$ and $\eta$ are functions of $\rho$ and $z$ given by
\begin{equation}\label{landeta}
\begin{array}{rcl@{\hspace{10mm}}rcl}
L &=& (l_{+} + l_{-})/2 & \eta &=& (l_{+} - l_{-})/2  \\[1.5mm] 
l_{+} &=& \sqrt{\rho^{2} + (z+A)^{2}}& 
l_{-} &=& \sqrt{\rho^{2} + (z-A)^{2}} \, .
\end{array}
\end{equation}
In these coordinates, the horizon $H$ is the line segment $\mid \!  z
\!\mid \leq A$ on the $\rho =0$ axis.  Both $\bar{\psi}$ and
$\bar{\gamma}$ are regular everywhere except in the limit $\rho
\rightarrow 0$ (for $\mid \!  z \!\mid \leq A$) where they diverge
logarithmically.  This divergence is nothing more than the usual
coordinate singularity at the event horizon.  To see this explicitly,
one can perform the following coordinate transformation from the Weyl
coordinates $(t,\rho,z,\phi)$ to the standard $(t,r,\theta,\phi)$
coordinates for Schwarzschild spacetime,
\begin{equation}\label{schwarzcoord}
\begin{array}{rcl@{\hspace{10mm}}rcl}
    r &=& L + A &
    \cos\theta &=& (l_{+} - l_{-})/2A \\[1.5mm]
    z &=& L\,\cos\theta &
    \rho^{2} &=& (L^{2} - A^{2})\sin^{2}\theta \, .
\end{array}
\end{equation}
The metric takes the usual Schwarzschild form with the line segment
$H$ mapped to the event horizon $r = 2A$.  Since the Weyl coordinates
are valid only in a region where the Killing vector $t^a$ is timelike,
these coordinates cover only the exterior region of the Schwarzschild
solution.  However, one can use the standard methods (see, for
example, chapter 6 of \cite{rmw}) to extend the spacetime inside the
horizon and show that there is no singularity at $H$.

We shall now construct the distorted black hole solutions.  Recall
that the field equation (\ref{psi}) is linear in the function
$\bar{\psi}$.  This enables us to add any harmonic function
$\bar{\psi}_D$ to the Schwarzschild potential $\bar{\psi}_S$ and
obtain a new solution $\bar{\psi} = \bar{\psi}_S + \bar{\psi}_D$ to
(\ref{psi}); this new solution may be considered to be a distorted
version of the Schwarzschild solution.  Furthermore, if $\bar{\psi}_D$
is regular everywhere, the new potential $\bar{\psi}$ will also be
logarithmically divergent at the line segment $H$ and thus the
location of the horizon will be unchanged.  However, since
$\bar{\psi}_{D}$ satisfies Laplace's equation and is regular at the
horizon, it will not tend to zero at infinity.  

Given this $\bar{\psi}$, one can easily integrate the remaining field
equations (\ref{gamma}) to obtain the second metric function $\bar{\gamma}$.
Although the field equations for $\bar{\gamma}$ are not linear, it is still
helpful to express $\bar{\gamma}$ as the Schwarzschild function plus a
distortion $\bar{\gamma}_{D}$ so that
\begin{equation}\label{decomp}
    \bar{\psi} = \bar{\psi}_{S} + \bar{\psi}_{D} \quad \mathrm{and} \quad
    \bar{\gamma} = \bar{\gamma}_{S} + \bar{\gamma}_{D} \, .
\end{equation}
Recall that the function $\bar{\gamma}$ must vanish on the axis due to
our boundary conditions.  In the present case, the axis consists of
two disjoint pieces so we must ensure that it is consistent to set
$\bar{\gamma}_{D} = 0$ on both portions of the axis.  It turns out
that this condition will also place a restriction on $\bar{\psi}_{D}$.  To
see this, substitute (\ref{decomp}) into the field equations
(\ref{gamma}) to obtain equations for $\bar{\gamma}_{D}$.  We shall
only be interested in the z-derivative of $\bar{\gamma}_{D}$,
\begin{equation}\label{psidz}
    \bar{\gamma}_{D,z} = 2\rho(\bar{\psi}_{S,\rho}\bar{\psi}_{D,z} +
    \bar{\psi}_{D,\rho}\bar{\psi}_{S,z} +
    \bar{\psi}_{D,\rho}\bar{\psi}_{D,z})  \, .\\
\end{equation}
Integrate this equation along a line parallel to and near $H$.  Only
the first term on the right hand side will contribute and, since
$\bar{\psi}_{S,\rho} = 1/\rho + \mathcal{O}(1)$, it follows that
$\bar{\psi}_{D}$ must have the same value at the two ends of $H$. 
Therefore, in order for $\bar{\gamma}$ to vanish on both disconnected
sections of the axis, we require that $\bar{\psi}_{D}$ takes the same
value $\bar{u}$ at both ends of the line segment $H$ .  Similarly, by
integrating (\ref{psidz}) from one end of $H$ to an arbitrary point on
$H$, it
follows that%
\footnote{Here, and throughout the paper, $\hateq$ will denote equality 
only at H.} 
\begin{equation}\label{ubar}
    \bar{\gamma}_{D} \hateq 2 \bar{\psi}_{D} -2\bar{u} \, .
\end{equation}
This result allows us to show that the metric is non-singular at the
horizon.  We shall prove this for the more general case of distorted
charged black holes in the next section.

Finally, the metric can be expressed in Schwarzschild coordinates using
(\ref{schwarzcoord}) and the decomposition of $\bar{\psi}$ and 
$\bar{\gamma}$ given in (\ref{decomp}).  It takes the form:
\begin{equation}\label{distschwarz}
    ds^{2} = -e^{2\bar{\psi}_{D}}\left(1-\frac{2A}{r}\right)\, dt^{2}+
    e^{2(\bar{\gamma}_{D}-\bar{\psi}_{D})}
    \left[ \frac{1}{(1-2A/r)}\, dr^{2} + r^{2}\, d\theta^{2} \right] +
    e^{-2\bar{\psi}_{D}}r^{2}\sin^{2}\theta \, d\phi^{2} \, .
\end{equation}
These are the distorted black hole solutions obtained in 
\cite{gh,ms}.  It has been shown in \cite{gh} that the metric can be 
analytically continued through the horizon.

Let us now consider the behaviour of these solutions at infinity. 
First of all, if we assume $\bar{\psi}_{D}$ is harmonic everywhere and
is not a constant, then it must diverge at infinity (if
$\bar{\psi}_{D}$ is constant, by changing coordinates we can set it to
zero).  Therefore these solutions are not asymptotically flat. 
However, it is possible to find asymptotically flat extensions if we
require that $\bar{\psi}_{D}$ is harmonic only in a neighborhood of
the horizon and extend $\bar{\psi}_{D}$ and $\bar{\gamma}_{D}$ so that
they tend to zero at infinity.  In the intervening region we assume
that $\bar{\psi}_{D}$ is not harmonic i.e. the vacuum Einstein
equations are not satisfied in this region.  In other words, there are
some matter fields present in this region.  This matter can be
interpreted as causing the distortion of the black hole.  Moreover, if
the matter satisfies the strong energy condition, it follows that
$\bar{u} \leq 0$.

To demonstrate that $\bar{u}$ is non-positive, let us consider
Einstein's equations projected in the $t$ direction, as in \cite{gh}:
\begin{equation}\label{tab}
    \Delta \bar{\psi}_{D}  \equiv \bar{\psi}_{D,\rho\rho} +
    \rho^{-1}\bar{\psi}_{D,\rho} + \bar{\psi}_{D,zz} =
    8\pi e^{2(\bar{\psi}_{D}-2\bar{\gamma}_{D})}(T_{ab} -
    \frac{1}{2}Tg_{ab})t^{a}t^{b} \, .
\end{equation}
If the matter satisfies the strong energy condition, the right hand
side of equation (\ref{tab}) is necessarily non-negative.  Since the
Laplacian is a negative operator and we have taken $\bar{\psi}_{D}
\rightarrow 0$ at infinity, $\bar{\psi}_{D}$ must be non-positive
everywhere in spacetime. In particular, this implies that $\bar{u} \leq
0$.

In the undistorted case, $\bar{\psi}_{D} = \bar{\gamma}_{D} = 0$, $A$ 
represents the ADM mass or equivalently the Komar mass evaluated at 
infinity.  However, this is \textit{not} the case with non-zero 
distortion.  The Komar mass of the spacetime as measured at infinity
is not equal to $A$. This can be demonstrated easily by using the
expression for the Komar mass:
\begin{equation}\label{komarmass}
M^{Komar}_{\infty} - M^{Komar}_{H} = 2\int_{\Sigma} \left(T_{ab} -
\frac{1}{2}Tg_{ab}\right) n^{a}\xi^{b} dV
\end{equation}
where $\Sigma$ is a constant time spacelike hypersurface, $n$ is the
unit timelike normal to $\Sigma$ and $\xi = \partial/\partial t$ is
the Killing vector which is unit timelike at infinity.  It is
straightforward to evaluate the Komar integral at the horizon and one
obtains $M^{Komar}_{H} = A$.  We already know from the strong energy
condition that the integrand on the right hand side of
(\ref{komarmass}) is non-negative everywhere due to the strong energy
condition.  We can also argue that it must be positive somewhere: If
it is identically zero, then from (\ref{tab}) $\bar{\psi}_D$ satisfies
the vacuum equations everywhere.  As argued previously, the solution
cannot be vacuum everywhere \textit{and} be asymptotically flat at
infinity.  Therefore, the right hand side of (\ref{komarmass}) is
necessarily positive.  Hence, we see that $M^{Komar}_{\infty} >
M^{Komar}_{H} = A$.  This can be understood quite easily by
considering the $t t$ component of the distorted Schwarzschild metric
(\ref{distschwarz}): $g_{tt} = e^{2\bar{\psi}_{D}}
\left(1-\frac{2A}{r}\right)$.  The Komar mass is equal to $1/(2r^2)$
times the derivative of $g_{tt}$ evaluated at infinity.  We obtain
\begin{equation}\label{kommass}
    M^{Komar}_{\infty} = A + \lim_{r\rightarrow \infty}(r^2 \partial
\bar{\psi}_{D}/\partial r) \, .
\end{equation}
However, from (\ref{tab}), we see that $\bar{\psi}_{D}$ satisfies a
Laplace equation with non-zero sources.  Therefore, it must tend to
zero as $\mathcal{O}(1/r)$ at infinity and the second term will contribute to
the Komar mass.  This is consistent with the previous conclusion that
$M^{Komar}_{\infty} > A$.

\subsection{Distorted Reissner--Nordstr\"om Solutions}
\label{s3.2}

The vacuum-electrovac correspondence discussed in section \ref{s2}
allows us to transform any given solution of the vacuum Weyl equations
to a solution of the electrovac Weyl equations.  It should therefore be
possible to find the electrovac solutions corresponding to the
distorted Schwarzschild black holes.  As we shall see, these new
solutions represent distorted Reissner--Nordstr\"om solutions.

The Reissner--Nordstr\"om spacetime is a static, axisymmetric solution
to the Einstein--Maxwell equations.  Therefore it can be cast in Weyl
form with the metric functions $\psi$ and $\gamma$ given by
\begin{equation}
\psi_{RN} = \frac{1}{2}\ln \left(\frac{L^{2} - A^{2}}{(L+M)^{2}}\right) 
\qquad {\textrm{and}} \qquad 
\gamma_{RN} = \frac{1}{2}
   \ln \left(\frac{L^{2} -A^{2}}{L^{2} - \eta^{2}}\right) \, .
\end{equation}
$L$ and $\eta$ are functions of $\rho$ and $z$ which have the same
functional form as in the Schwarzschild case (\ref{landeta}) and now
$A$ is given in terms of the mass $M$ and charge $Q$ of the black hole
as $A:= \sqrt{M^{2} - Q^{2}}$. The electric potential is given by
\begin{equation}
\Phi = \frac{Q}{L+M} \, .
\end{equation}
In the limit $Q \rightarrow 0$, the electric potential becomes zero
and we recover the Schwarzschild solution of mass $M$ in Weyl
coordinates.  In order to recast this metric in standard
Reissner--Nordstr\"om coordinates $(t,r,\theta,\phi)$, perform the
following coordinate transformation
\begin{equation} \label{rntrans}
\begin{array}{rcl@{\hspace{5mm}}rcl}
r &=& L + M & \cos\theta &=& (l_{+} - l_{-})/2A \\[1.5mm] z &=&
L\,\cos\theta & \rho^{2} &=& (L^{2} - A^{2})\sin^{2}\theta \, .
\end{array}
\end{equation}
As in the Schwarzschild solution discussed in section \ref{s3.1}, the
horizon $H$ is a line segment on the $z-$axis with $\mid \!  z \!\mid
\leq A$.  The Weyl coordinates cover only the region outside the event
horizon at $r = R \equiv M + A$.  It is clear that the spacetime can
be extended through the horizon in the standard way.

The functional relation (\ref{eqforphi}) between $\Phi$ and $\psi$ is
satisfied in Reissner--Nordstr\"om spacetime with $C=M/Q$ and $v=0$. 
Therefore, it must be possible to find a vacuum Weyl solution which
can be transformed into the Reissner--Nordstr\"om solution. 
Intuitively one might expect this vacuum solution to be the
Schwarzschild solution.  This is indeed the case: the Schwarzschild
solution with mass $A=\sqrt{M^2 -Q^2}$ is mapped to the
Reissner--Nordstr\"om solution with mass $M$ and charge $Q$ under the
transformation with $C=M/Q$ and $v=0$.  In particular,
\begin{equation} \label{rnbar}
    \bar{\psi}_{RN} = \frac{1}{2}\ln\left(\frac{L-A}{L+A}\right)
        \, , \qquad
    \bar{\gamma}_{RN} = \frac{1}{2}\ln\left(\frac{L^{2} - A^{2}}
        {L^{2}-\eta^{2}}\right) \, , \qquad
    C = \frac{M}{Q} \qquad \mathrm{and} \qquad
    v = 0 \, .
\end{equation}
Interestingly, even if we choose the constant $C$ to be less than unity
so that $\bar{\psi}$ is imaginary, the metric functions $\psi$ and
$\gamma$ are still real and the metric describes a Reissner--Nordstr\"om
solution with $Q > M$. We refer the reader to \cite{gha} for more
details.

Since the equation for $\bar{\psi}$ is linear, we can ``distort'' the
Reissner--Nordstr\"om solution just as we distorted the Schwarzschild
solution.  In particular, we can add to $\bar{\psi}_{RN}$ any regular
solution $\bar{\psi}_{D}$ of the Laplace equation (\ref{psi}).  We can
then solve for $\bar{\gamma}$ and decompose it into the
Reissner--Nordstr\"om function $\bar{\gamma}_{RN}$ and a distortion
$\bar{\gamma}_{D}$.  Thus, as in the distorted Schwarzschild case, we
obtain
\begin{equation}\label{psigamd}
\bar{\psi} = \bar{\psi}_{RN} + \bar{\psi}_{D} \qquad \rm{and} \qquad
\bar{\gamma} = \bar{\gamma}_{RN} + \bar{\gamma}_{D} \, .
\end{equation}
We can now use the vacuum-electrovac transformation discussed in
(\ref{s2}) to find the electrovac solution corresponding to
$\bar{\psi}$ and $\bar{\gamma}$.  The transformation involves the two
parameters $C$ and $v$ and we need to specify them.  If
$\bar{\psi}_{D}$ is identically zero, then we want the transformed
solution to be the Reissner--Nordstr\"om solution.  This means that we
must choose $C=M/Q$ as before.%
\footnote{We could also choose $C$ based on the criteria that any
solution where $\bar{\psi}_{D}$ is everywhere constant (but not
necessarily zero) should be transformed to the Reissner--Nordstr\"om
solution.  This leads to $C = (M\cosh\bar{u} + A\sinh\bar{u})/Q$. 
However, most of the results including the first law are unchanged by
this choice and for the sake of simplicity, we have chosen $C=M/Q$.}
As for the parameter $v$, recall that we chose it to be zero in
section \ref{s2} because we required asymptotic flatness.  However, we
can no longer impose this condition since we have required
$\bar{\psi}_{D}$ to be harmonic and regular everywhere which means
that it diverges at infinity.  Thus, there is no longer any reason to
choose $v=0$ and we shall just leave it as a free parameter in the
transformation.  (Asymptotic flatness will again be achieved by
putting additional matter fields away from the black hole.)

Using (\ref{eqforpsi}) with $C = M/Q$ and $v$ a free parameter, the
metric function $\psi$ is given (in Reissner--Nordstr\"om coordinates)
by
\begin{equation}\label{psid}
    e^{2\psi} =
    \dr e^{2\psi_{D}}
    \quad \mathrm{where} \quad
    e^{-\psi_{D}} := e^{-v}\left[\cosh\bar{\psi}_{D}
       - \left(1-\frac{Q^{2}}{Mr}\right)\frac{M}{A}\sinh\bar{\psi}_{D}
       \right] \, .
\end{equation}
Here, we have defined $\dr$ as
\begin{equation}\label{delta}
    \dr = \left(1 - \frac{2M}{r} + \frac{Q^{2}}{r^{2}}\right) \, .
\end{equation}
One might wonder if there is an expression similar to (\ref{ubar})
relating the values of $\psi_{D}$ and $\gamma_{D}$ at the horizon.  It
is not obvious that the relationship $\bar{\gamma}_{D} \hateq
\bar{\psi_{D}}-2\bar{u}$ generalizes to the unbarred quantities. 
However, we do know that $\gamma = \bar{\gamma}$ everywhere.  At the
horizon, where $r = M+A$, the relationship (\ref{psid}) simplifies so
that $\psi_{D} = \bar{\psi}_{D} + v$.  Therefore,
\begin{equation}\label{gampsihor}
    \gamma_{D} \hateq 2\psi_{D} - 2(\bar{u} + v)
\end{equation}
at the horizon of a distorted charged black hole. 

Let us now express the distorted Reissner--Nordstr\"om spacetimes in
terms of the familiar $(t,r,\theta,\phi)$ coordinates (\ref{rntrans}). 
The metric is found by substituting the forms of $\psi$ and $\gamma$
given by (\ref{psigamd}), (\ref{psid}) into the Weyl metric.  We 
obtain:
\begin{equation}\label{rnmetric}
    ds^{2} = - \dr e^{2\psi_{D}}dt^{2}
    + \frac{e^{2(\gamma_{D} - \psi_{D})}}{\dr}dr^{2}
    + e^{2(\gamma_{D} - \psi_{D})}r^{2}d\theta^{2}
    + e^{-2\psi_{D}}r^{2}\sin^{2}\theta\, d\phi^{2}\, .
\end{equation}
To specify the solution fully, we must also give the form of the
electric potential.  This is found by solving (\ref{eqforphi}) to
obtain
\begin{equation}\label{phirn}
   \Phi = e^{v}\left(\frac{M}{Q} - \sqrt{\frac{M^2}{Q^2} -1 +
   \dr e^{2\psi_{D}-2v}} \,\right)  \, .
\end{equation}
In order to show that these are indeed distorted charged black holes,
it is necessary to demonstrate that the metric is regular at the
horizon $\dr = 0$.  The easiest way to analyze this matter is to pass
to ingoing Eddington-Finkelstein coordinates.  Introduce a new
coordinate $w$ satisfying $dw = dt + e^{-2(\bar{u} + v)}\,dr/\dr$. 
Thus, $w$ is similar to the standard ingoing Eddington-Finkelstein
coordinate.  The metric can be expressed in $(w,r,\theta, \phi)$
coordinates as
\begin{eqnarray}\label{rnef}
    ds^{2} &=&
    -\dr e^{2\psi_{D}}\,dw^{2}
    + 2e^{2\psi_{D}-2(\bar{u} + v)} \,dw\, dr
    + e^{2\psi_{D}}\left(\frac{e^{2\gamma_{D} - 4\psi_{D}}-
    e^{-4(\bar{u} + v)}}
    {\dr}\right) \, dr^{2} \nonumber \\
    && \qquad\qquad + e^{2(\gamma_{D} - \psi_{D})}r^{2} \, d\theta^{2}
    + e^{-2\psi_{D}}r^{2}\sin^{2}\theta \, d\phi^2 \, .
\end{eqnarray}
At first it appears that the coefficient of the $dr^2$ term becomes
infinite at the horizon.  However, (\ref{gampsihor}) guarantees that
both the numerator and denominator vanish at the horizon.  One can
then expand both expressions in powers of $\rho^{2}$ to show that the
coefficient remains finite (but generically non-zero) at the horizon.

Since the spacetime is not asymptotically flat, it is not possible to
define the concept of an event horizon (since this requires a notion
of null infinity).  However, there are several reasons which suggest
these solutions contain black holes.  Firstly, the surface $r = R := M
+ A$ is a Killing horizon of the Killing vector $\xi \propto
\frac{\partial}{\partial w}$.  Secondly, the more general notion of an
isolated horizon \cite{abf,afk} is also applicable here.  Clearly, all
the isolated horizon conditions are satisfied here since this is a
Killing horizon.  These are two local justifications for calling this
a black hole solution.  Finally, it is not difficult to show as in
\cite{gh} that this solution can be extended to be asymptotically
flat, in which case the horizon will truly be the event horizon of a
black hole.  To do so, assume the solution presented above is valid
only in a neighbourhood of the horizon.  Outside this region, $\psi$,
$\gamma$ and $\Phi$ can be extended arbitrarily so that they tend to
zero at spatial infinity.  In the intervening region, the electrovac
equations (\ref{einsmax}) will not be satisfied.  This indicates the
presence of non-electromagnetic matter in the region.  Physically, one
can think of this matter as causing the distortion of the black hole. 
As in the uncharged case, if the matter satisfies the strong energy
condition it follows that $\bar{u} + v \leq 0$.

The Komar mass at the horizon for the charged solution is $A$ and the
Komar mass at infinity is now $M$ plus the `$1/r$ part of $\psi_{D}$'. 
Physically, we expect $\psi_{D}$ to fall off as $1/r$ and be negative
whence the Komar mass at infinity must be strictly greater than $M$. 
Intuitively, this is reasonable because, as in the
Reissner--Nordstr\"om solution, the difference $(M-A)$ should be the
energy in the electromagnetic field whereas the $1/r$ part of
$\psi_{D}$ gives the energy in the additional matter fields.  However,
it is not clear how to make this statement more precise because we
have extended the electromagnetic potential in an essentially
arbitrary fashion and the energy in the electromagnetic field may not
be exactly $(M-A)$.

\subsection{Properties}
\label{s3.3}

Let us now turn our attention to some of the properties of the metric
given above.  In particular, we shall focus attention on the horizon $r
= R := (M+A)$.  The metric on a cross section of the horizon is given
by
\begin{equation}\label{2metric}
    ds^2 = e^{2(\gamma_{D} - \psi_{D})}R^{2} \, d\theta^{2}
    + e^{-2\psi_{D}}R^{2}\sin^{2}\theta \, d\phi^2 \, .
\end{equation}
The area of a cross section of the horizon is easily found, making use
of (\ref{gampsihor}), to be
\begin{equation}\label{area}
    a_{{\mathrm{H}}} \hateq 4\pi R^{2}e^{-2(\bar{u} + v)}
\end{equation}
whence the area radius of the horizon $\rh$ is given by $\rh := R
e^{-(\bar{u} + v)}$.  It is already clear from (\ref{2metric})
that the cross sections of the horizon are indeed distorted.  To make
this more explicit, we can calculate the curvature of any 2-sphere
cross section of the horizon,
\begin{equation}\label{curv}
    {}^{2}R \=\frac{e^{-2\psi_{D}}}{\rh^{2}}
    \left(1-2(\psi_{D,\theta})^{2} + 3\cot\theta \, \psi_{D,\theta}
    + \psi_{D,\theta\theta} \right).
\end{equation}
Clearly, this is not constant on the 2-sphere.  However, in the limit
$\psi_{D}, \gamma_{D} \rightarrow 0$, the curvature tends to $1/\rh^2$
as expected for a round 2-sphere.

The electromagnetic field strength can be calculated from the
potential (\ref{eqforphi}) to be
\begin{equation}
   \emF = \frac{e^{2\psi_{D}-v} }{(A^2/Q^2 + \dr e^{2\psi_D -2v})^{1/2}}\,
    dt \wedge \left[
   \dr (\psi_{D,\theta} d\theta + \psi_{D,r}dr) +
   \left(\frac{M}{r^2}-\frac{Q^2}{r^3}\right) dr \right]\,
\end{equation}
which reduces to $\emF \= (Q/R^{2})e^{2\psi_{D}-v} \, dr \wedge dt$ at
the horizon.  Dualizing this equation, we see that
\begin{equation}\label{starf}
  \dual{\emF} \= \frac{Q e^{-v}}{\rh^{2}} \, {}^{2}\epsilon
\end{equation}
where ${}^{2}\epsilon = \rh^{2}\sin\theta \, d\theta \wedge d\phi $ is
the volume form on any 2-sphere cross section of the horizon.  Thus,
equation (\ref{starf}) implies that the ``effective charge density''
of the horizon is uniform.  In some sense, even though the horizon is
distorted, we see that the electric field at the horizon has been
distorted in exactly the same manner as the geometry.  Integrating
(\ref{starf}) over the horizon, we find that the electric charge of
the black hole is given by
\begin{equation}\label{charge}
    Q_{\mathrm{H}} \hateq Q e^{-v} \, .
\end{equation}

As for any isolated horizon, the solution we have presented here is of
type II at the horizon. This is because the static Killing vector $\xi$
is a repeated principal null direction of the Weyl Tensor at the
horizon (see e.g. \cite{afk}). Thus in a null tetrad adapted to the
horizon,%
\footnote{$\l$ is chosen parallel to the degenerate direction of
the horizon, $m$ and $\bar{m}$ are tangent to the horizon and
transverse to $\l$ and $n$ is transverse to the horizon.}%
the components of the Weyl tensor $\Psi_{0}$ and $\Psi_{1}$ vanish at
the horizon.  This implies that at the horizon, $\Psi_{2}$ is
invariant under null rotations about $\l$ and is given by
\begin{equation}\label{psi2}
\Psi_{2} \= \frac{\qh^{2}}{2\rh^{4}} -
\frac{e^{-2\psi_{D}}}{2\rh^{2}}\left\{1-2(\psi_{D,\theta})^{2} + 3\cot
\theta \, \psi_{D,\theta} + \psi_{D,\theta\theta} \right\} \, .
\end{equation}
Similarly, we can explicitly show that $\Phi_{00}$, $\Phi_{10}$ and
$\Phi_{20}$ all vanish.  The only relevant component of the Ricci
tensor at the horizon is
\begin{equation}
    \Phi_{11} := \frac{1}{4} R_{ab}(\l^a n^b + m^a \bar{m}^b) =
    \frac{\qh^2}{2\rh^4}.
\end{equation}

\section{Black Hole Thermodynamics}
\label{s4}

In this section we will discuss the zeroth and first laws of black
hole mechanics for the distorted, charged black holes.  We shall see
that the zeroth law holds, i.e. the surface gravity of the black hole
is constant.  However, the form of the first law is sensitive to the
choice of normalization of the timelike Killing field.  We shall
discuss two such choices: a local choice arising from isolated horizon
considerations and a global choice of normalization at infinity.  Both
choices lead to a first law, but we shall see that the two versions
take very different forms.

\subsection{Zeroth Law}

Let us first turn our attention to the surface gravity, $\kappa$.  The
surface gravity of a Killing horizon is typically defined as
\begin{equation}
    \xi^a \nabla_a \xi^b = \kappa \xi^b \, ,
\end{equation}
where $\xi$ is the horizon generating Killing vector.  However, this
definition does not give the surface gravity uniquely --- there is a
freedom to rescale $\xi$ by a constant which results in a constant
rescaling of $\kappa$.  This freedom is usually fixed by appealing to
infinity.  In the class of spacetimes under consideration, the
extension of the solution to infinity is by no means unique and will
contain matter fields which we do not model.  Hence, there is no
natural normalization so we will only define $\xi$, and hence
$\kappa$, up to a constant rescaling.  This freedom does not affect
the zeroth law: if the surface gravity is constant for one choice of
$\xi$ it is constant for every $\tilde{\xi}$ related to the original
one by constant rescaling.  Thus, we need only show that $\kappa_{0}$
corresponding to the choice $\xi_{0} = \left( \frac{\partial}{\partial
w} \right)$ is constant.  The surface gravity is given by
\begin{equation}\label{kappa}
    \kappa_{0} \hateq \frac{e^{2(\bar{u} + v)}}{2 R}\left(1 -
    \frac{Q^2}{R^2} \right) \hateq \frac{e^{(\bar{u} +v)}}{2 \rh}
    \left(1 - \frac{\qh^2 e^{-2\bar{u}}}{\rh^2} \right).
\end{equation}
The surface gravity $\kappa_{0}$ is clearly constant over the horizon. 
The zeroth law of black hole mechanics is satisfied.

Let us now turn our attention the value of the electromagnetic
potential at the horizon.  We would also expect this to be constant on
the horizon of a black hole.  Substituting $r = M + A$ into
(\ref{phirn}), it follows that
\begin{equation}\label{potential}
    \ph \hateq \frac{Qe^{v}}{R} \hateq \frac{\qh e^{v - \bar{u}}}{\rh} \, .
\end{equation}
Thus, as expected from general considerations, the electric potential
is constant on the horizon.  The zeroth law of black hole mechanics
and ``the electromagnetic version of the zeroth law'' hold at the
horizon of the distorted charged black holes.

\subsection{Local First Law}
\label{s4.1}

Let us now consider a version of the first law for these black holes
arising from the isolated horizon framework.  In order to make the
arguments in this section, we assume that the spacetime has been
extended to be asymptotically flat by the addition of matter fields
which admit a Hamiltonian description.  Furthermore, we assume that
these matter fields do not have support in a neighbourhood of the
black hole, i.e. in that neighbourhood, the Einstein--Maxwell
equations hold.  With these assumptions, the isolated horizon 
results \cite{afk} are directly applicable to distorted charged black 
holes.

In the isolated horizon framework one attempts to find a Hamiltonian
describing evolution along a vector field $t^a$.  We assume that the
magnitude of $t^{a}$ is unit at infinity and at the horizon $t^a
\propto \left(\frac{\partial}{\partial w}\right)^{a}$; this is
appropriate since the horizon is non-rotating.  The proportionality
factor must be a constant in a given spacetime but may vary in phase
space, for example it may depend upon the area and charge of the
horizon and the constants $\bar{u}$ and $v$; in other words, we allow
$t^{a}$ to be a live vector field.  Now one can ask the following
question: is there a Hamiltonian $H^{t}$ describing time evolution
along $t^{a}$?  The answer is in the affirmative if and only if there
exists a function $E_{H}^{t}$ only of $\ah$ and $\qh$ such that the
first law holds:
\begin{equation}\label{ihfirstlaw}
    \delta E_{H}^{t} = \frac{1}{8\pi} \kappa_{t} \delta a_{H} +
    \Phi_{t} \delta Q_{H} \, .
\end{equation}
It turns out that $E_{H}^{t}$ is the horizon surface term in the
Hamiltonian $H^{t}$ and is therefore interpreted as the horizon energy
associated with time translation along $t^{a}$.  Furthermore the
surface gravity $\kappa_{t}$ and electric potential $\Phi_{t}$
associated with $t^{a}$ are also functions only of $\ah$ and $\qh$, in
particular they do not depend upon the distortion parameters $\bar{u}$
and $v$.  We shall call $t^{a}$ an admissible vector field if
(\ref{ihfirstlaw}) is satisfied: Every admissible $t^{a}$ gives rise
to a first law.

It is natural to ask whether any choice of time evolution vector field
$t^{a}$ is `canonical' in a suitable sense.  For the black holes
described in this paper, there is indeed a preferred choice
$t_{0}^{a}$ which is normalized appropriately at the horizon.  As
mentioned above, the first law (\ref{ihfirstlaw}) implies that both
$\kappa_{t}$ and $\Phi_{t}$ can be functions only of $\ah$ and $\qh$. 
However, there exists a two parameter family of Reissner--Nordstr\"om
solutions labelled by $\ah$ and $\qh$.  These solutions are in the
phase space of distorted charged black holes; they simply correspond
to the absence of any distorting matter.  In order that the surface
gravity and electric potential of the horizon take their standard
values in Reissner--Nordstr\"om spacetime, we must necessarily choose
$t_{0}^{a}$ such that
\begin{equation}\label{khph}
    \kappa_{t_{0}} \hateq \frac{1}{2 \rh} \left(1-\frac{\qh^{2}}{\rh^{2}} \right)
    \quad \mathrm{and} \quad
    \Phi_{t_{0}} \hateq \frac{\qh}{\rh} \, .
\end{equation}
This not only guarantees that $\kappa$ and $\Phi$ take their usual
values on Reissner--Nordstr\"om spacetimes, but also fixes their
functional form for \textit{all} distorted charged black holes:  There
is a unique choice $t^{a}_{0}$ of time evolution at the horizon
for which $\kappa_{t_{0}}$ and $\Phi_{t_{0}}$ are given by
(\ref{khph}).  The energy associated with this time evolution will be
denoted by $E_{H}^{t_{0}}$.  By definition, $E_{H}^{t_{0}}$ satisfies
the first law with $\kappa_{t_{0}}$ and $\Phi_{t_{0}}$ given in
(\ref{khph}).  Furthermore, if we require that the energy tends to
zero as both $a_{H}$ and $Q_{H}$ tend to zero, it follows that
\begin{equation}\label{smarr}
   E_{H}^{t_{0}} = \frac{1}{4 \pi} \kappa_{t_{0}} \ah 
   + \Phi_{t_{0}}\qh \, .
\end{equation}
Therefore, the isolated horizon framework provides a canonical
definition of the energy of a distorted black hole.

\subsection{Global First Law}
\label{s4.2}

There is a second form of the first law for distorted charged black
holes.  Here we wish to consider normalizations appropriate to an
observer at infinity, much as Geroch and Hartle did for the uncharged
case \cite{gh}.  They obtained a first law $\delta M = \frac{1}{8 \pi}
\kappa_{H}\delta a_{H} + M_{H}^{Komar}\delta \bar{u}$, where
$M_{H}^{Komar}$ is the Komar mass of the horizon and $\kappa_{H}$ is
the surface gravity associated with the Killing field normalized to
unity at infinity.  The parameter $\bar{u}$ was interpreted as the
potential due to external matter and the extra term in the first law
as work done on the black hole by this matter.  The parameter $M$ is
interpreted as ``the mass of the black hole alone as measured at
infinity'' (in this case it also happens to be equal to
$M_{H}^{Komar}$).  More specifically, it is the first term in the
expression for the Komar mass at infinity (\ref{kommass}), but is
certainly not the mass of the entire spacetime --- there will be a
second contribution from the distorting matter.

We shall now describe how to obtain a similar version of the first law
for distorted charged black holes.  In order to do this, we must first
give a prescription by which the spacetime is to be extended to
infinity.  This extension is essentially arbitrary and involves
unknown matter fields.  First, given the metric obtained previously
(\ref{rnmetric}),
\begin{equation}
    ds^{2} = - \dr e^{2\psi_{D}}dt^{2}
    + \frac{e^{2(\gamma_{D} - \psi_{D})}}{\dr}dr^{2}
    + e^{2(\gamma_{D} - \psi_{D})}r^{2}d\theta^{2}
    + e^{-2\psi_{D}}r^{2}\sin^{2}\theta\, d\phi^{2}\, ,
\end{equation}
we extend this to be asymptotically flat as before by requiring
$\psi_{D}$, $\gamma_{D}$ and $\Phi \rightarrow 0$ as $r \rightarrow
\infty$.  This will not be possible if the Einstein--Maxwell equations
are satisfied \textit{everywhere} in spacetime.  Therefore, we assume
that the spacetime satisfies the Einstein--Maxwell equations in a
neighbourhood of the horizon only.  Outside this region, we allow
other kinds of matter which we shall not model precisely.  However,
this matter is parametrized by the constants $\bar{u}$ and $v$ which
affect the values of the various parameters of the horizon:
\begin{equation}\label{horpara}
\begin{array}{rcl@{\hspace{5mm}}rcl}
    a_{H} &=& 4\pi R^{2}e^{-2(\bar{u} + v)} &
    Q_{H} &=& Qe^{-v}\\[1.5mm]
    \kappa_{H} &=& \frac{e^{2(\bar{u} + v)}}{2 R}
       \left(1- \frac{Q^2}{R^2} \right)
    &\ph &=& \frac{Qe^{v}}{R}\, .
\end{array}
\end{equation}
Here, $R = (M+A)$ and $A^{2} = M^{2} -Q^{2}$.  

In order to obtain a version of the first law, one must first decide
how to define the mass of the black hole.  One obvious choice is the
Komar mass of the horizon, $M_{H}^{Komar} = A \equiv (1/4\pi) \kh
\ah$.  However, this is not a good definition of black hole mass
because it does not give the correct answer for the undistorted
Reissner--Nordstr\"om solutions.  The Komar mass does not include
contributions from the electromagnetic ``hair'', but is just the
gravitational mass of the black hole.  The other natural choice is
$M$, the parameter appearing in the metric (\ref{rnef}).  Why should
this be interpreted as the mass of the black hole?  Firstly, from the
definition of $A$, it follows that
\[M = \sqrt{(M^{Komar}_{H})^{2} +Q^{2}}\]
so it clearly contains a contribution from the electromagnetic field. 
Furthermore, it is straightforward to show that $M$ satisfies a Smarr
formula: $M = (1/4\pi) \kh \ah + \ph \qh$.  Therefore, one may
interpret it as the mass of the horizon including contributions from
both the gravitational and electromagnetic fields.  Thus, it seems
reasonable to interpret $M$ as the total mass of the black hole while
$M^{Komar}_{H}$ represents only the gravitational contribution to the
mass.

Now we have decided that the mass of the black hole is $M$, one can
simply vary the above expressions for $\ah$ and $\qh$ to obtain the
algebraic identity:
\begin{equation}\label{global1law}
    \delta M = \frac{1}{8 \pi} \kappa_{H}\delta a_{H} +
    \ph \delta \qh + \sqrt{(M^{Komar}_{H})^{2} +Q^{2}}\,\, \delta v +
    M^{Komar}_{H} \delta\bar{u} \, .
\end{equation}
This is the first law applicable for an observer at infinity.  As
usual, all terms on the right hand side are evaluated at the horizon
(although the normalizations of $t$ and $\Phi$ are determined at
infinity).  This form of the first law (\ref{global1law}) is similar
to that obtained in the uncharged case.  In the charged case there are
now two extra work terms which must be interpreted.  Before doing so,
let us discuss the parameters $\bar{u}$ and $v$.  First of all, note
that the quantity $\bar{u}$ comes directly from the distorted
Schwarzschild solution and therefore it tells us how the
\textit{uncharged} part of the external matter affects the hole. 
Another way to see this is to notice that in (\ref{horpara}),
$\bar{u}$ only affects the values of gravitational parameters at the
horizon, $\kh$ and $\ah$, but not the electromagnetic ones.  On the
other hand, $v$ affects both the gravitational and electromagnetic
parameters.  Therefore, we can think of $v$ as describing the
effective potential due to some charged matter present in the
spacetime.

Let us now return to the first law.  The first two terms on the right
hand side of (\ref{global1law}) are the usual ones describing changes
in mass of the black hole due to changes in area and charge.  The
third term is a work term explaining how the mass of the black hole
changes as we change $v$.  We have argued above that $v$ is a
potential due to some charged matter in the spacetime.  Therefore, it
seems reasonable that $v$ couples to the total mass
$\sqrt{(M^{Komar}_{H})^{2} +Q^{2}}$ which contains contributions from
both gravitational and electromagnetic fields.  In contrast, $\bar{u}$
represents the uncharged matter in the spacetime, so it only couples
to the gravitational part of the mass, $M^{Komar}_{H}$.  Thus, the
$M^{Komar}_{H}\delta \bar{u}$ term can be interpreted as the
gravitational work done on the black hole by variations in the
uncharged external matter while the $\sqrt{(M^{Komar}_{H})^{2} +Q^{2}}
\, \delta v$ is the work done by charged external matter.

These considerations are, of course, only heuristic.  More
importantly, although we have argued that $M$ should be interpreted as
the total mass of the black holes since it satisfies a Smarr formula,
its physical significance is in fact not clear.  It is certainly not
the ADM mass of spacetime.  Due to the arguments given at the end of
section \ref{s3.1}, the ADM mass must also contain a contribution from
${\psi}_{D}$.  In the distorted Schwarzschild case, $M$ is the Komar
mass at the horizon but even this is not true in the charged case.  As
in \cite{gh}, we may perhaps think of $M$ as ``the mass of the black
hole alone as measured by an observer at infinity''; i.e. it is the
quantity obtained by ignoring all contributions pertaining to
$\psi_{D}$ in the expression for the ADM mass.

\section{Conclusion}

In this paper we have obtained the first family of distorted charged
black hole solutions.  These solutions generalize the distorted
Schwarzschild solutions studied in \cite{gh,ms} to Einstein--Maxwell
theory.  In order to obtain black holes with non-zero charge, we have
used the vacuum-electrovac correspondence found in \cite{w,gha} to
transform the distorted Schwarzschild solutions to distorted
Reissner--Nordstr\"om solutions.  These solutions are regular at the
horizon, but are not asymptotically flat unless one includes
additional matter fields in the exterior portion of spacetime.  The
black holes are clearly distorted since the 2-curvature of the horizon
is not constant.

We have also studied the zeroth and first laws of black hole
mechanics.  The surface gravity and electric potential are both
constant on the horizon as expected.  However, when considering the
first law, one must introduce a normalization of the vector field
generating the horizon and also an electromagnetic gauge choice.  We
have presented two different choices.  The first is motivated by
isolated horizons and leads to the standard form of the first law. 
Furthermore, it is natural to normalize the vector field generating
time translation such that the surface gravity and electric potential
have the same functional form as in Reissner--Nordstr\"om spacetimes. 
This framework also leads to a canonical definition of the energy of
the black hole.

The second alternative is to normalize the Killing vector and electric
potential at infinity.  This leads to formulae for the surface gravity
and electric potential which depend upon two extra parameters,
$\bar{u}$ and $v$.  These can be thought of as describing in some
sense the amounts of uncharged and charged matter in spacetime.  We
have also obtained a first law tailored to this choice.  As one might
expect, the first law contains extra terms involving $\bar{u}$ and $v$
which can be thought of as work terms involving the uncharged and
charged matter respectively.  It is important to emphasize that the
interpretations given for the global first law are all somewhat
heuristic.  In particular the parameter $M$ which is to be interpreted
as the mass of the black hole is not equal to the ADM mass of
spacetime.  Therefore, we feel that the isolated horizon framework
provides a clearer interpretation of the first law for these black
holes.

\section*{Acknowledgments}

We are most grateful to Abhay Ashtekar and Chris Beetle for numerous 
discussions and to Josh Willis for carefully proof-reading the paper. 
This work was supported in part by the NSF grants PHY94-07194,
PHY95-14240, INT97-22514 and by the Eberly research funds of Penn
State.  SF was supported in part by a Braddock Fellowship.

\end{document}